\documentclass[preprint,12pt]{elsarticle}
\usepackage{amsmath,amsfonts}
\usepackage{algorithmic}
\usepackage{algorithm}
\usepackage{array}
\usepackage[caption=false,font=normalsize,labelfont=sf,textfont=sf]{subfig}
\usepackage{textcomp}
\usepackage{verbatim}
\usepackage{rotating}
\usepackage{makecell}
\usepackage{comment}
\usepackage{hyperref}
\usepackage{booktabs}
\usepackage{siunitx}
\usepackage{pifont}
\usepackage{tabularx}
\usepackage{graphicx}
\usepackage{pdflscape}
\usepackage[table]{xcolor}
\newcolumntype{Y}{>{\centering\arraybackslash}X}
\newcolumntype{P}[1]{>{\centering\arraybackslash}p{#1}}

\begin{document}

\begin{frontmatter}

\title{High-Resolution 3D Micro-CT Database of Transcranial Magnetic Stimulation (TMS) Coils for More Accurate Modelling, Technology Benchmarking, and Progress in Coil Design}

\author{Maryam Farahmandrad}

\author{Max Koehler}
\author{Mahdi Bayati}
\author{Denis Goldnik}
\author{Sergiy Antonyuk}
\author{and Stephan M. Goetz\corref{cor}}

\cortext[cor]{Send correspondence to smg.}

\begin{abstract}
Accurate modelling of transcranial magnetic stimulation (TMS) coils requires coil-specific geometric and electrical parameters. Yet detailed manufacturer models are rarely available, and internals may be widely available to companies in the field but not to researchers. The development of new coils should ideally build on all existing prior art. This work presents a technology benchmark and modelling-ready dataset for seventeen widely used commercial TMS coils across multiple manufacturers and geometries. High-resolution micro-CT (49.5~µm resolution, material and density contrast) serves to reconstruct winding and casing structures and to extract parameters directly usable in numerical models, including loop dimensions, conductor cross-section, turn count, winding height, and winding-to-surface (casing) thickness. Complementary electrical measurements provide spectral impedance (series resistance and inductance) scans from 20~Hz to 200~kHz, together with dc resistance. In addition to baseline geometry and impedance, the study documents practical construction features relevant to safety and usability with detailed insights into their internal construction (e.g., cooling architectures and embedded sensing). The resulting benchmark dataset supports coil-specific electric-field simulations, reduces modelling uncertainty associated with generic assumptions (e.g., fixed coil–scalp spacing), and provides a quantitative reference for designing and evaluating new coil technologies against established commercial solutions. As a simulation-readiness example, we generated a calibrated CT-derived MCB70 line-current model for SimNIBS and evaluated its effect on predicted electric-field metrics. Compared with the SimNIBS MCB70 model, the CT-derived model produced measurable differences under identical placement conditions. We provide the dataset online.
\end{abstract}


\end{frontmatter}

\section{Introduction}
Transcranial magnetic stimulation (TMS) is a non-invasive method to probe and modulate brain circuits \cite{rossini2015non}. TMS uses strong, brief current pulses to induce  a spatial electric field profile through a stimulation coil to write signals into neural circuits or to modify endogenous neuronal activity \cite{lu2017comparison}. The locus of activation in the brain is approximately in the area where the induced electric field is maximal \cite{krings1997stereotactic}. The design of the TMS coil significantly affects this location and efficacy of stimulation delivery in the brain \cite{drakaki2022database, thielscher2004electric}. In order to provide a map of direct neural activation by TMS, the induced electric field distributions generated by different coil types have been characterised by theoretical calculations \cite{tofts1990distribution, branston1991analysis, esselle2002neural, jalinous1991technical, ravazzani1996magnetic, roth1990theoretical, ruohonen1998transcranial, zimmermann1996slinky, eaton1992electric}, numerical simulation models \cite{leo1992computer, saypol1991theoretical,toschi2009transcranial, sekino2002comparison, kowalski2002current, toschi2008reconstruction, sekino2004fem, salinas2007detailed}, and measurements of the electric currents induced in phantoms \cite{maccabee1993magnetic, tay1989measurement} or \textit{in vivo} \cite{wagner2004intracranial, sances1991mapping}. However, only a few commercial coils have been modelled in more detailed computational studies, and analytical studies typically employ highly idealised circular and/or figure-of-eight coil geometries \cite{salinas2007detailed, thielscher2004electric, lontis2006focality, epstein2002iron}.

 The geometry and structure of the TMS coil determine the strength and spatial distribution of the induced electric field, which determines the focality of stimulation and the depth at which neurons can be stimulated \cite{ueno2021figure, deng2014coil}. However, despite the wide range of coil designs available, detailed information on key characteristics remains insufficiently documented in the literature \cite{mancino2020obtaining}. Knowledge of the characterisation of specific coils and how they compare to other coils is valuable for selecting the optimal coil for a given application or to accurately model and predict stimulation outcomes, as well as the development of novel coils.
 For most older coils, although once widely used, their winding pattern and the technology are poorly characterised and models are missing as manufacturers closed the business or stopped manufacturing for other reasons. As they have been neither characterised nor modelled, available results cannot be fully interpreted, and valuable coil technology is forgotten.

Different manufacturers have created a variety of TMS coil designs, each incorporating unique technological features to influence the stimulation properties \cite{drakaki2022database}. Different coil geometries,  which include parameters such as coil size, shape, and number of turns, provide distinct advantages to target particular regions of the brain. The initial TMS system used a circular coil, which induces a non-focal ring-shaped electric field under the coil perimeter \cite{bickford1965non,polson1982stimulation,deng2017development}. The desire to overcome the limitations of circular coils in terms of focality has led to the development of the figure-of-eight coil, also known as the butterfly coil, which consists of two adjacent loops and produces a more concentrated electric field at their intersection \cite{ueno1990focal,ueno1990functional}. The figure-of-eight coil enables much more focal brain stimulation, and it can generate a higher electric field at the targeted centre point compared to areas just a few centimetres away. Thus, it rapidly became the standard for cortical mapping and targeted stimulation in both research and clinical settings \cite{ueno2021figure}.

Practically all coils for focal brain stimulation to date use some form of figure-of-eight design. In these coils, the two circles are arranged side by side or overlap. Most coils are furthermore flat. Angled or curved figure-of-eight coils in contrast to flat ones can improve the efficiency \cite{koehler2024closed,koponen2015minimum,thielscher2004electric}. The curved or angled surface can reduce the required pulse energy and improve efficiency but typically complicates handling for many operators and perfectly tangential alignment on the round head \cite{koehler2024coil}. 
Non-concentric figure-of-eight windings in contrast to previous (and still dominant) concentric loops spread out the outer parts of the loops and concentrate the windings as well as the induced electric field in the centre. This innovation could improve efficiency and focality \cite{knaulein1996minimization,koponen2015minimum}.

Other coil configurations have been developed to achieve specific stimulation goals. Racetrack coils feature an elongated winding shape that covers a broader area than a standard circular coil, typically for neuromuscular and other peripheral stimulation \cite{koponen2018multi, goetz2011comparison, goetz2022prediction,kammer2003physikalische}.  In contrast, the cone coil tilts the winding to conform to scalp curvature, to increase the depth of penetration \cite{lontis2006focality}. Leaf-type coils concentrate magnetic stimulation under a small petal loop to maximise focality. Leaf designs may help with precise stimulation of small peripheral or cranial nerve targets \cite{roth1994vitro}.  

An approach to increase efficiency is the incorporation of ferromagnetic cores into the coil. An iron core can increase the magnetic field for a given current and reduce power consumption and heat production \cite{lorenzen1992practical,al1998novel,epstein2002iron}. Although cores can increase efficiency and reduce cooling needs, all known magnetic materials, without any exception, increase coil weight \cite{koehler2025magnetic}.

So-called H coils distribute the winding across the entire frontal part of the head \cite{roth2002coil, roth2007three, huang2009consensus, harel2011h}. The design includes a base portion that runs largely tangential to the scalp, whereas the return portions typically remain farther from the head, where they produce greater stray flux associated with energy need and loss \cite{koehler2024simplified}.
In contrast to focal stimulation with figure-of-eight designs, H coils offer more widespread stimulation. The smaller dependence on highly precise and accurate placement may explain their high efficacy in clinical applications \cite{filipvcic2019efficacy,levkovitz2015efficacy,oreardon2007efficacy,lefaucheur2010image,goetz2021neuronavigation}.
H coils are currently only offered by one supplier so that their availability remains limited despite their apparently high performance. Limited knowledge about the internals also seems to hamper further development, both commercially and academically. First analyses indicate considerable potential for energy and complexity improvements \cite{koehler2024simplified}.

\begin{figure}[!t]
    \centering
    \includegraphics[width=0.3\columnwidth]{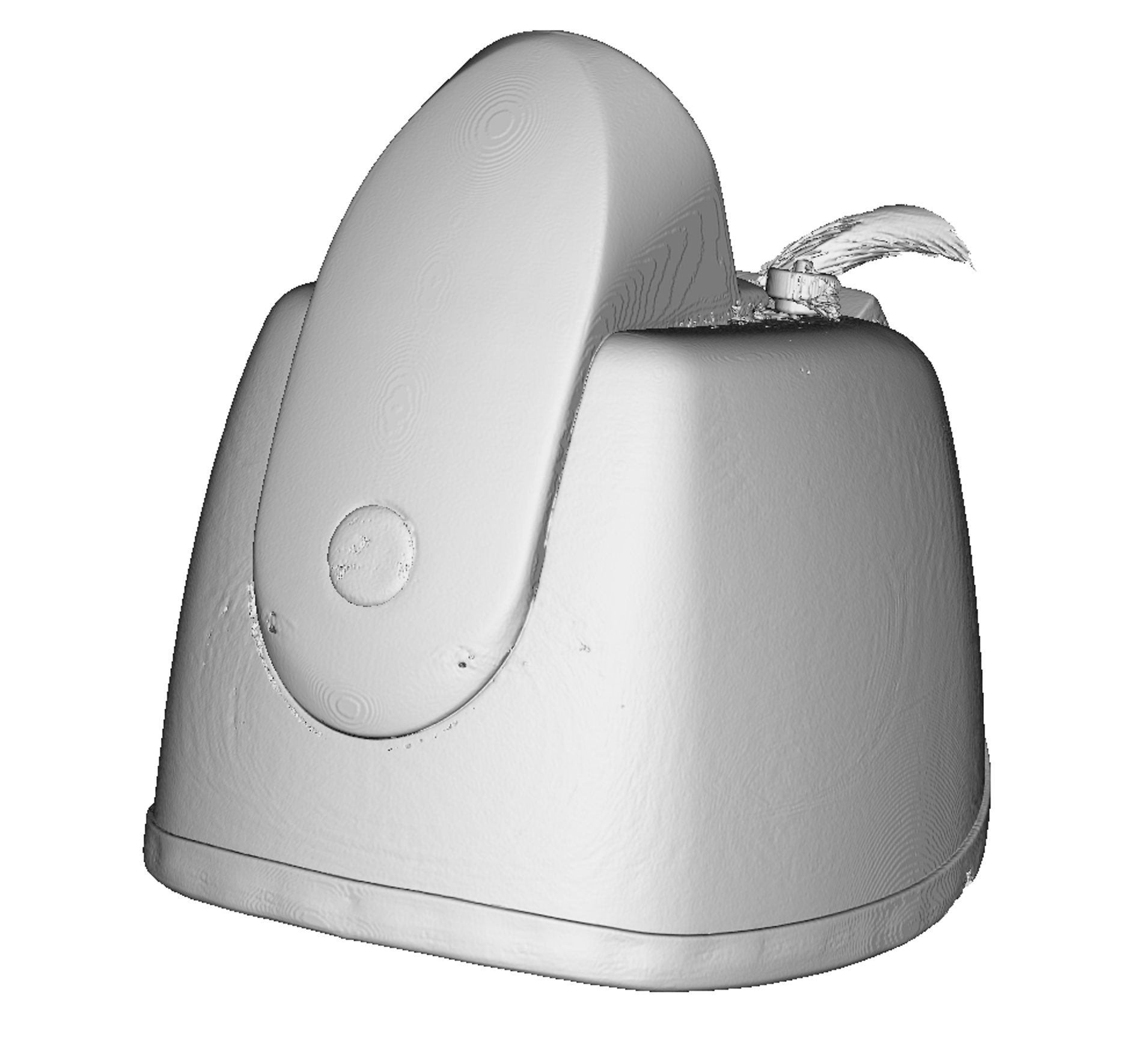}
    \includegraphics[width=0.3\columnwidth]{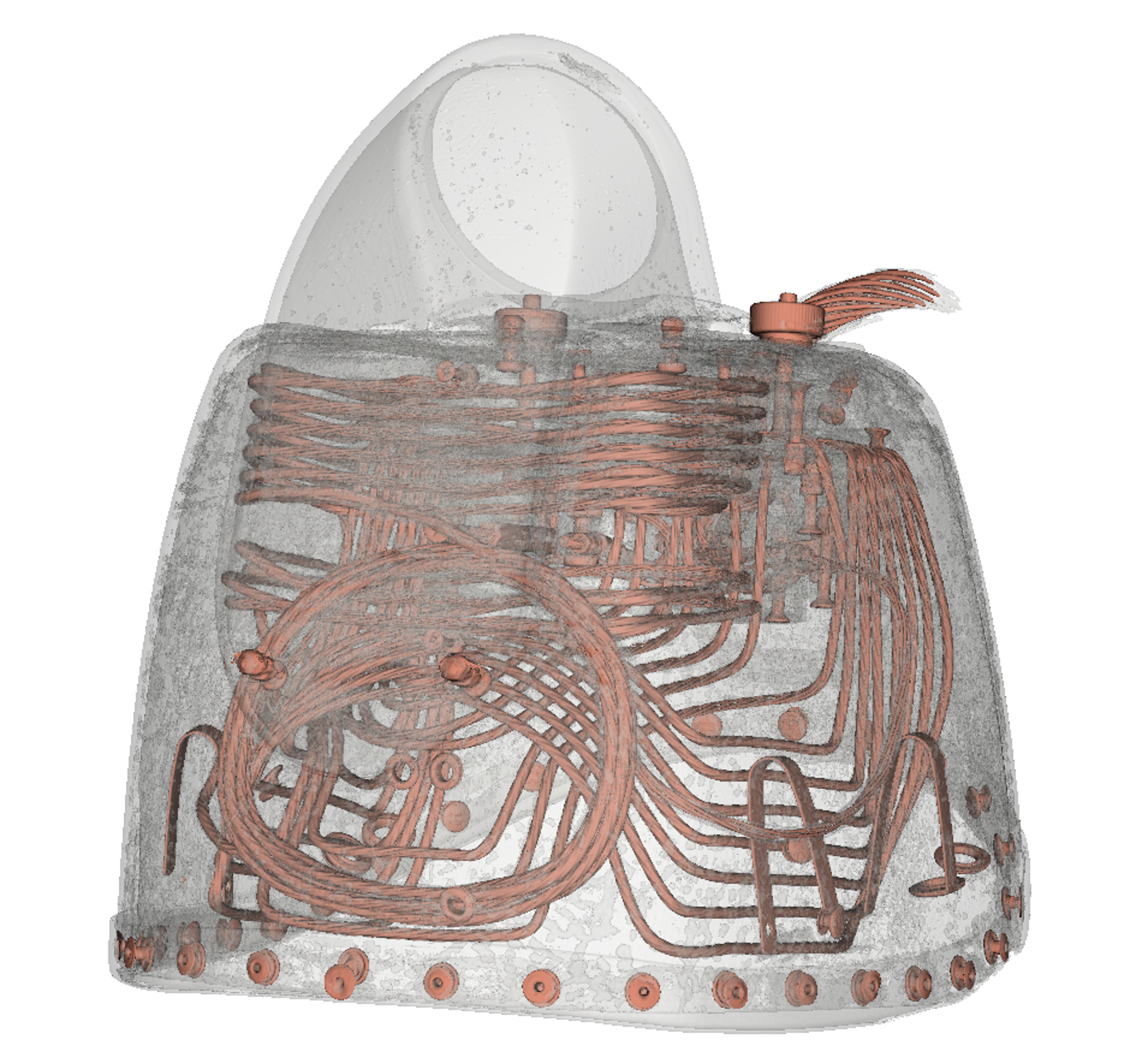} 
    \includegraphics[width=0.3\columnwidth]{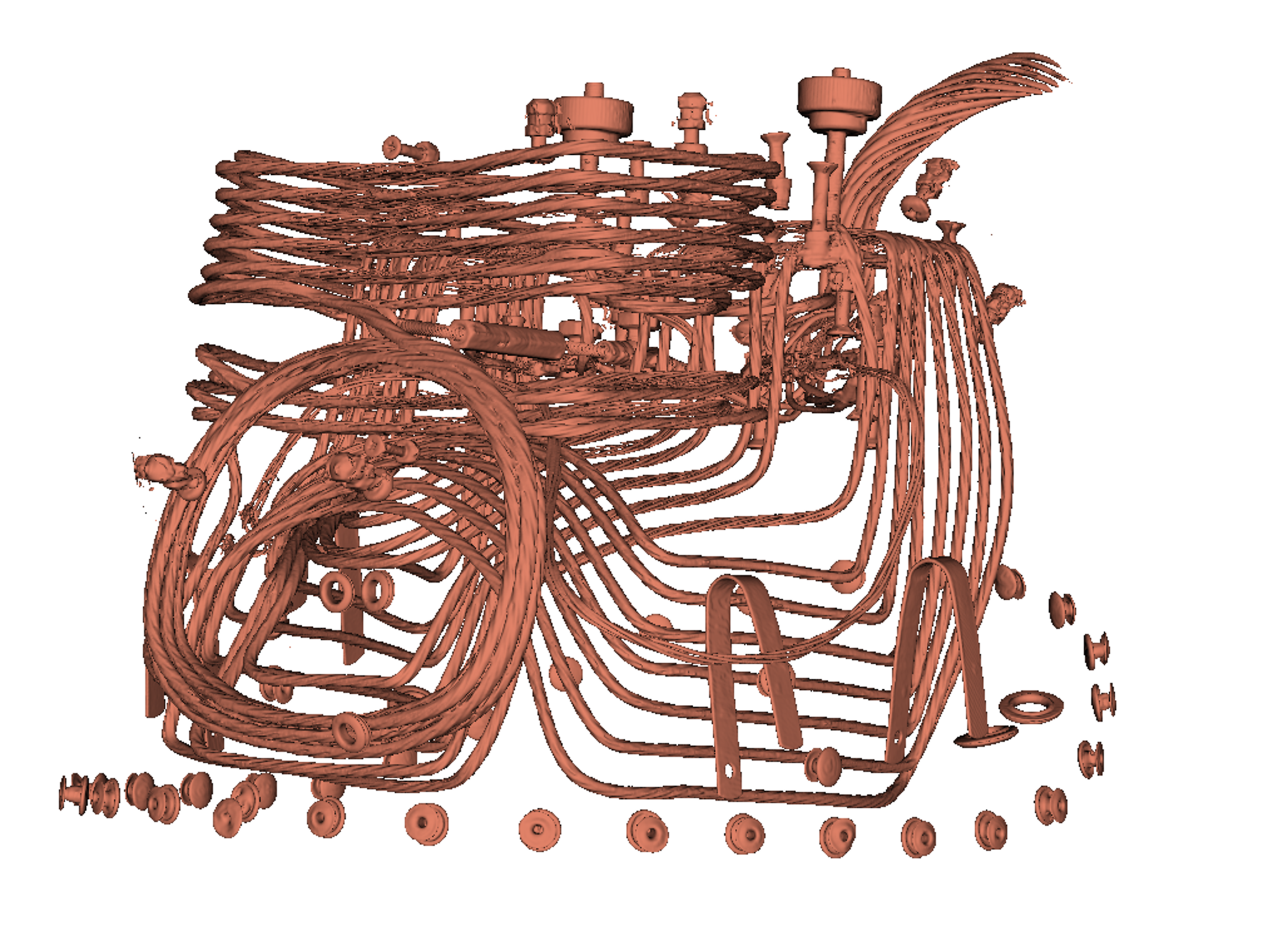}
    \caption{3D visualisations of segmented micro-CT scans of an H1 coil: complete coil including the plastic housing and internal structure of the conductor geometry.}
    \label{3D_Picture}
\end{figure}

The geometry of a TMS coil, which includes the outer and inner loop diameters, the amount of loop overlap in figure-of-eight designs, and the coil-to-scalp casing gap, controls the spread and focus of the induced electric field \cite{wang2023investigating}. Larger outer diameters tend to increase the overall stimulation area, whereas smaller inner diameters can concentrate the magnetic field and make it stronger at the target \cite{gomez2018design}. In figure-of-eight coils, greater loop overlap increases focality by concentrating the magnetic field at their convergence \cite{ueno2021figure}.
The casing gap, defined as the distance between the windings and the scalp, affects field strength, the threshold, and efficiency. A smaller gap increases the induced field for a given current, but it can also increase heat transfer to the scalp and the risk of high-voltage breakdown. For this reason, coil design must balance stimulation strength with patient safety \cite{stokes2005simple, caulfield2024mitigating}.

The conductor and winding parameters of a TMS coil, such as the number of turns and the wire cross-section, drive practical trade-offs among field strength, efficiency, and heat build-up. A higher turn count increases the inductance (up to the squared turn number if they are tightly coupled) and trades off pulse voltage against current \cite{babbs2014compact}. 

Modern TMS coils rely not only on geometry but also on conductor technology to improve performance and safety. Some TMS coils use solid copper conductors, whereas others use litz wire composed of numerous individually insulated strands arranged in bundled, twisted, braided, or otherwise transposed configurations to reduce high-frequency losses and heating \cite{weyh2005marked,elfahem2024advanced}.

Despite the wide range of available TMS coils and reported technical features, coil-specific information required for accurate computational modelling is often unavailable or inconsistently reported. Accurate electric field simulation depends on precise knowledge of coil dimensions, winding layout, and relevant material/interface properties, which are not publicly available for every model \cite{gomez2020conditions}. To fill this gap, we provide a technology benchmark and modelling-ready dataset for seventeen widely used coils from multiple manufacturers. High-resolution micro-CT serves to reconstruct winding and housing structures and to extract parameters usable in numerical models (e.g., loop dimensions, conductor cross-section, turn count, winding height, and winding-to-surface thickness), complemented by electrical characterisation including resistance, inductance (5 kHz), and broadband impedance signatures. The parameters or winding paths can feed existing software packages to generate accurate simulation models \cite{koehler2023tms}. The resulting dataset enables reproducible coil-specific simulations and provides a quantitative reference to support modelling and the design and evaluation of improved next-generation coil technologies.

This paper presents four main outcomes:
\begin{itemize}
    \item[--] A micro-CT-derived 3D geometry database providing coil-specific dimensions for numerical electric-field modelling.
    \item[--] Standardised electrical characterisation of the frequency-dependent impedance, including resistance and inductance.
    \item[--] Discovery and documentation of construction features (e.g., cooling architecture, wire properties, and embedded sensing) relevant to safety, usability, and modelling assumptions.
    \item[--] A documented simulation example demonstrating how CT-derived MCB\-70 winding geometry can be reconstructed as a SimNIBS-compatible line-current model and evaluated against the SimNIBS MCB70 model under matched head-model, placement, distance, conductivity, and stimulation settings.
    
\end{itemize}

\section{Materials and Methods}

This study provides a  characterisation  of seventeen commercial TMS coils from seven manufacturers (MagVenture, Dantec, Cadwell, Magstim, Rogue Research, Neuronetics, and BrainsWay) aimed at modelling and technology development. We selected these coils to reflect a wide range of models used in clinical and research settings today. The workflow combines high-resolution micro-computed tomography (micro-CT) with targeted electrical measurements to extract geometric and construction parameters that are critical for accurate coil modelling. These include loop dimensions, casing thickness below the windings, winding
height, conductor cross-section and layering, and special internal features. We designed all procedures to yield quantities that can be directly used in numerical simulation environments.

\subsection{Overall Characterisation Protocol}

In order to obtain quantities that can readily feed numerical models, we followed a standard protocol. First, we used a high-resolution micro-CT scanner (Werth Tomoscope L, Gießen, Germany) to acquire a high-resolution three-dimensional dataset of the winding region and case. From these volumes, we segmented the conductor and casing to form the basis for all geometric analyses such as loop dimensions,  casing thickness underneath the windings, winding height, conductor cross-section, and number of layers.
Figure~\ref{3D_Picture} illustrates a representative 3D rendering generated from the segmented micro-CT volume. 

Subsequently, we measured the electrical properties of each coil with a precision LCR meter (Hameg Instruments HM8118, Frankfurt, Germany) in four-wire Kelvin configuration. We acquired the complex broadband impedance, \(Z(f)\), across the specified frequency range. From these measurements, we extracted the frequency-dependent real part of the impedance, \(\Re\{Z(f)\}\), and the effective inductance, \(\Im\{Z(f)\}/\omega\), where \(\omega = 2\pi f\). We further extracted the series resistance and series inductance of each coil at 5 kHz.

\subsection{ Data Post-Processing}
The geometric and electrical quantities shown in Tables~\ref{Coil_Physical_Properties}–\ref{Coil_Electrical_Properties} were obtained by post-processing the electrical measurements and CT data. 

In order to describe the coil-level geometry (Table~\ref{Coil_Physical_Properties}), the outer and inner loop diameters were measured directly from the micro-CT data. For coils with circular loops, the outer diameter defines the distance between the outer edges of the windings, and the inner diameter is the distance between the inner edges of the windings. We defined the coil–casing gap (Table~\ref{Coil_Physical_Properties}) as the distance between the outer surface of the winding pack and the outer surface of the coil on the stimulation side.

For the simulation-oriented comparison, we also calculated the winding-centre-to-surface distance. We calculated this distance as the patient-side casing thickness plus one half of the winding height at the same in-plane location. This value estimates the distance from the patient-side coil surface to the centre of the winding pack, which defines the effective current-path position in a line-current coil representation.

Table~\ref{Coil_CrossSection} provides a summary of conductor-level parameters. The cross-sectional micro-CT data determined the kind of conductor (solid bar or litz wire). Additionally, the visible winding passes in the CT slices indicated the number of turns. CT slices orthogonal to the wire trajectory yielded the  conductor cross-sectional area. The conductor lengths reported in Table~\ref{Coil_CrossSection} refer to the internal coil windings.

Table~\ref{Coil_Electrical_Properties} lists the electrical parameters at 5~kHz. For each coil, electrical characterisation in series mode determined the series resistance and series inductance at 5~kHz. We also measured the broadband impedance response between 20 Hz and 200 kHz. The complex impedance data were then post-processed to obtain \(\Re\{Z(f)\}\) and \(\Im\{Z(f)\}/\omega\), which describe the frequency-dependent resistive and effective inductive behaviour of each coil.

\subsection{Coil-Model Generation}

To demonstrate the simulation applicability  of the extracted geometry, we rebuilt the MagVenture MCB70 coil as a SimNIBS-compatible line-current model. We generated the model from the segmented micro-CT winding geometry rather than from scalar table values alone. The segmented CT winding geometry was first converted into turn-wise centreline paths for the line-current approximation. Since the MCB70 coil had a CT-derived turn count of $(2\times10)$, the SimNIBS-compatible model was constructed with 20 closed line-current elements, with each element representing the centreline of one winding turn. Ten elements represented the left lobe and ten represented the right lobe. Opposite current directions were assigned to the two lobes to reproduce the figure-of-eight coil configuration.

The effective current path was placed near the centre of the winding pack. We therefore calculated the line-current depth as the patient-side casing thickness plus one half of the measured winding height. For the MCB70 coil, this calculation gave a calibrated line-current depth of 6.3 mm from the outer contact surface towards the winding pack. We rebuilt the directed centrelines at this calibrated depth and converted them into a SimNIBS-compatible \texttt{.tcd} coil file using the SimNIBS line-segment coil API. The rebuilt \texttt{.tcd} model contained 20 line-segment coil elements.

We compared the rebuilt CT-derived MCB70 model with the SimNIBS MCB70 model in SimNIBS 4.6.0 and the \texttt{m2m\_ernie} head model. For both simulations, we used the same target location, coil orientation, input scalp distance, conductivity setting, and stimulation rate. We set the target to C3, oriented the coil y-direction towards CP3, set the input scalp distance to 4.0 mm, and explicitly defined the stimulation rate as $\mathrm{d}I/\mathrm{d}t = 1.0\times10^{6}$ A/s. The final SimNIBS logs reported the same coil-placement matrix for both models and the same coil-to-cortex distance of 18.59 mm. Thus, the coil model representation formed the only intended difference between the two simulations: the rebuilt CT-derived line-current model versus the SimNIBS MCB70 model.

We extracted grey-matter electric-field percentiles and focality volumes from the SimNIBS outputs. The reported electric-field values correspond to $\mathrm{d}I/\mathrm{d}t = 1.0\times10^{6}$ A/s and can be linearly scaled to other stimulation rates.

\section{Results}

\subsection{Coil Geometries and Coil--Casing Gaps}
Table~\ref{Coil_Physical_Properties} summarises macroscopic geometry features of the seventeen coils, including loop size and shape, as well as the casing thickness between the windings and the stimulation surface. The set includes racetrack, round, cone, leaf, and helmet designs, along with eight figure-of-eight configurations from multiple manufacturers.

The inner loop diameters span roughly 24 mm to 112 mm, whereas the effective outer loop diameters range from about 46 mm in the compact Corticoil 5$\times$9 to about 160 mm along the long axis of the RT120 racetrack coil. This wide spread reflects different targeting goals, from large-coverage cortical stimulation to highly focal stimulation. Prior modelling studies suggest that larger loop diameters increase coverage and penetration depth at the cost of focality, whereas smaller loop diameters and curvatures concentrate the electric field near the largest winding density for higher focality. 

The BrainsWay H1 is a helmet-style distributed coil rather than a single-loop design, so its geometry cannot be characterised by a single loop diameter in the same way as conventional circular or figure-of-eight coils.

Table~\ref{Coil_Physical_Properties} also lists the overlap between the two loops for figure-of-eight coils. Increasing the overlap between wings of a figure-of-eight coil can increase focality and peak field strength at the intersection of the two loops. In this analysis, the MagVenture MCB70  shows a loop overlap of about 26~mm. On the other hand, the Cadwell Corticoil 5$\times$9, the Magstim Horizon, the four Rogue Research figure-of-eight coils, and the Neuronetics NeuroStar coil do not overlap.

An important additional geometric quantity is coil–casing gap, which is the thickness of the casing material between the outside surface of the winding pack and the contact surface. In addition to the winding height, the casing thickness on the patient side of the coil directly affects the effective coil-to-scalp distance.  An increase of the coil-to-scalp distance over a range of only a few millimetres due to wire height and/or casing thickness significantly lowers the peak electric field at the cortex. Our observations show that the internal space between the windings and the scalp-contacting surface in commercial coils ranges from 1.7 mm (Cadwell FocalPoint–9 cm leaf coil) to 7.5~mm (Horizon 3.0 E-z Cool coil). For the Magstim Horizon 3.0 E-z Cool coil, the winding-to-surface distance is approximately 7.5 mm, whereas the solid casing thickness is approximately 2.1 mm. The remaining distance includes internal space and structures associated with the coil's cooling design. 

Because the windings are deeper inside the housing, two coils with similar winding geometry and current waveform can have significantly different peak field strengths and motor thresholds. Coils with small coil--casing gaps and small loops (like the leaf designs) achieve higher peak fields for a given d$I$/d$t$ compared to coils with thicker housings and larger coil-to-scalp distances. 
In turn, their stimulation threshold is lower, and their efficiency is higher.

\begin{figure}[!t]
    \centering
    \includegraphics[width=0.75\columnwidth]{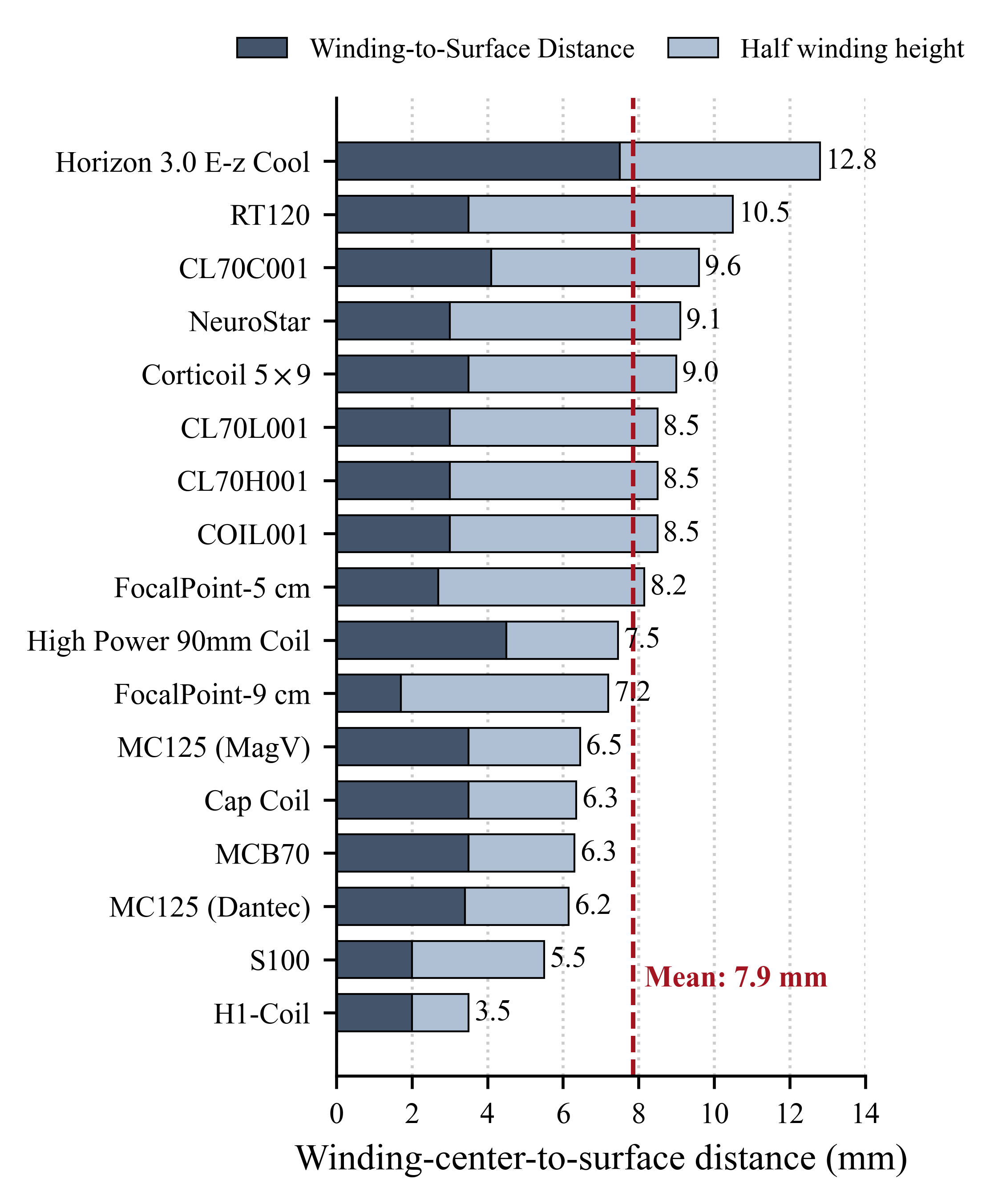} 
    \caption{Winding-centre-to-surface distance in the surveyed TMS coils.}
    \label{Winding-center-to-surface distance}
\end{figure}

Figure~\ref{Winding-center-to-surface distance} compares the winding-centre-to-surface distance across all seventeen coils. We calculated this distance by adding the measured patient-side casing thickness to half of the corresponding winding height (as a first approximation of the equivalent current position). This metric provides a more simulation-relevant estimate than casing thickness alone, because the effective current path in a line-current model lies near the centre of the winding pack rather than at the nearest winding boundary. The winding-centre-to-surface distance varied substantially across the surveyed coils, from 3.5 mm for the BrainsWay H1 coil to 12.8 mm for the Magstim Horizon 3.0 E-z Cool coil. This variation indicates that a single fixed winding-to-head spacing, such as a generic 5 mm insulation spacing \cite{deng2014coil}, may underestimate the effective current-source distance in several coils and thereby introduce coil-dependent modelling errors.

\begin{landscape}
\begin{figure*}[!t]
    \centering
    \includegraphics[width=1.42\textwidth]{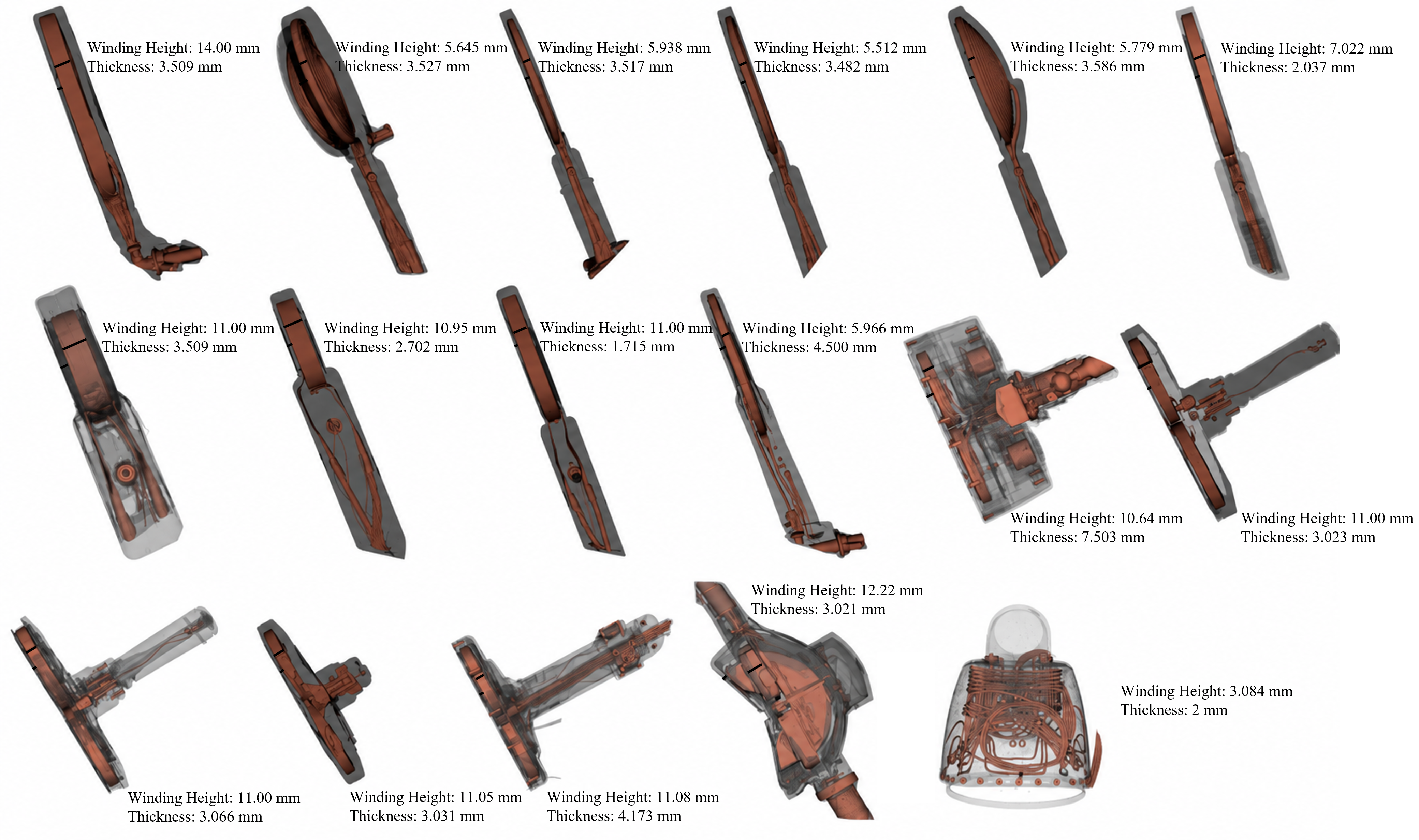} 
    \caption{Representative micro-CT orthogonal slice views showing the measurement markers used to extract coil--casing gap and winding height (measured at the same in-plane location). First row: MagVenture RT120, MagVenture MCB70, MagVenture MC125, Dantec MC125, Dantec Cap Coil, Dantec S100; second row: Cadwell Corticoil 5$\times$9, Cadwell FocalPoint--5 cm, Cadwell FocalPoint--9 cm, Magstim High Power 90 mm Coil, Magstim Horizon 3.0 E-z Cool, Rogue Research CL70H001; third row: Rogue Research CL70L001, Rogue Research COIL001, Rogue Research CL70C001, Neuronetics NeuroStar, BrainsWay H1 Coil.}
    \label{casing_winding_measurement}

\end{figure*}
\end{landscape}

Casing thickness was measured directly from the micro-CT volume for each coil. We recorded the corresponding winding height at the same in-plane measurement location. Figure~\ref{casing_winding_measurement} illustrates representative orthogonal slice views with the measurement annotations used for both quantities.

\begin{landscape}
\begin{table*}[!t]
\centering
\caption{Physical specifications of TMS coils.}
\label{Coil_Physical_Properties}
\footnotesize
\setlength{\tabcolsep}{4pt}
\renewcommand{\arraystretch}{1.12}

\begin{tabularx}{\linewidth}{@{} c l >{\raggedright\arraybackslash}X l c c c c @{}}
\toprule
\textbf{Nr} & \textbf{Manufacturer} & \textbf{Model} & \textbf{Coil geometry} &
\textbf{Outer loop Ø} & \textbf{Inner loop Ø} & \textbf{Loop overlap} & \textbf{Winding-to-Surface} \\
& & & & \textbf{(mm)} & \textbf{(mm)} & \textbf{(mm)} & \textbf{ Distance (mm)} \\
\midrule
1 & MagVenture & RT120 & Racetrack & 81.5$\times$160 & 31.5$\times$112&  -- & 3.5 \\ 
2 & MagVenture & MCB70 & Figure-of-Eight & 96.5 & 24.3 &  26 & 3.5 \\ 
3 & MagVenture & MC125 & Round & 114 & 28.1 &  -- & 3.5\\ 
4 & Dantec & MC125 & Round & 115.7 & 28.4 &  -- & 3.4\\ 
5 & Dantec & Cap Coil & Cone & 120 & 25 &  -- & 3.5 \\ 
6 & Dantec & S100 & Round & 97 & 69.7 &  -- & 2 \\ 
7 & Cadwell & Corticoil 5$\times$9 & Figure-of-Eight & 45.8 & 35 &  0 & 3.5 \\ 
8 & Cadwell & FocalPoint--5 cm & Leaf & 48.8 & 34.4 &  -- & 2.7 \\ 
9 & Cadwell & FocalPoint--9 cm & Leaf & 87.2 & 76.2 & -- & 1.7 \\ 
10 & Magstim & High Power 90 mm & Round & 123.5 & 68 &  -- & 4.5\\ 
11 & Magstim & Horizon 3.0 E-z Cool & Figure-of-Eight & 97 & 37 & 0  & 7.5 \\ 
12 & Rogue Research & CL70H001 & Figure-of-Eight & 90.8 & 49.3 &  0 & 3 \\ 
13 & Rogue Research & CL70L001 & Figure-of-Eight & 95.2 & 49.6 &  0 & 3\\ 
14 & Rogue Research & COIL001 & Figure-of-Eight & 95.1 & 30.4 & 0 & 3 \\ 
15 & Rogue Research & CL70C001 & Figure-of-Eight & 92 & 49.3 &  0 & 4.1\\ 
16 & Neuronetics & NeuroStar & Figure-of-Eight &48$\times$75 &36$\times$63 & 0 & 3 \\ 
17 & BrainsWay & H1 Coil & Helmet & -- & -- & -- &  2 \\ 
\bottomrule
\end{tabularx}
\end{table*}
\end{landscape}

\subsection{Conductor Technology and Winding Parameters}

Table~\ref{Coil_CrossSection} provides an overview of the conductor-level design of the coils, including conductor type, number of turns, and conductor cross-sectional area obtained from CT cross sections. The effective copper cross-section varies by an order of magnitude, from $2.5~\text{mm}^2$ in the compact Cadwell FocalPoint (5 cm leaf coil) to $28.9~\text{mm}^2$ in the large MagVenture RT120 racetrack coil. 

When combined with the electrical results in Table~\ref{Coil_Electrical_Properties}, these structural parameters help explain the observed inductance and resistance. Across the seventeen coils, the series resistance ranges from 13.9 to 82.8~m$\Omega$ and the series inductance from 7.5 to 26.2~µH at 5~kHz. Figure~\ref{coil_analysis}(A) shows a strong correlation between $L_s$ and $R_s$ (Pearson $r \approx 0.91$), so coils with higher inductance generally also exhibit higher series resistance. From an efficiency perspective, the MagVenture MCB70 figure-of-eight coil stands out with the highest inductance-to-resistance ratio (11.7~µH and 13.9~$\text{m}\Omega$), followed by RT120 and the Cap Coil. In contrast, the Cadwell Corticoil $5\times9$, the Cadwell FocalPoint–9 cm, and the Magstim High Power 90 mm Coil form a separate group with relatively high inductance and resistance.

For each coil, we computed the resistance per turn ($R_{\text{per turn}} = R_s/N$), where $N$ is the number of turns. Figure~\ref{coil_analysis}(B) plots $R_{\text{per turn}}$ against conductor cross-section. The data exhibit a negative correlation (Pearson $r \approx -0.6$) between conductor cross-sectional area and resistance per turn ($R_{\text{per turn}}$). Larger conductors typically produce lower resistance. The FocalPoint--9~cm coil is an example of a coil with a litz-wire conductor of small cross-sectional area (5.8~mm$^2$) and a relatively high $R_{\text{per turn}}$ (4.82~m$\Omega$/turn). 
The Magstim High Power 90 mm Coil has the highest resistance per turn in the dataset (5.9~m$\Omega$/turn at 10.6~mm$^2$), which shows that the combination of conductor length and cross-section may be unfavourable, particularly for high-power stimulation, e.g., neuromuscular \cite{madariaga2007magnetic, szecsi2010force}. The RT120, Cap Coil, and the two MC125 variations, on the other hand, represent a cluster with comparatively large cross-sections (15–29~mm$^2$) and lower $R_{\text{per turn}}$ (about 1.4–1.6~m$\Omega$/turn). The MCB70 coil has the lowest $R_{\text{per turn}}$ of all coils (0.70~m$\Omega$/turn at 14.7~mm$^2$). 

\begin{landscape}
\begin{figure*}[!t]
    \centering
    \includegraphics[width= 1.4\textwidth]{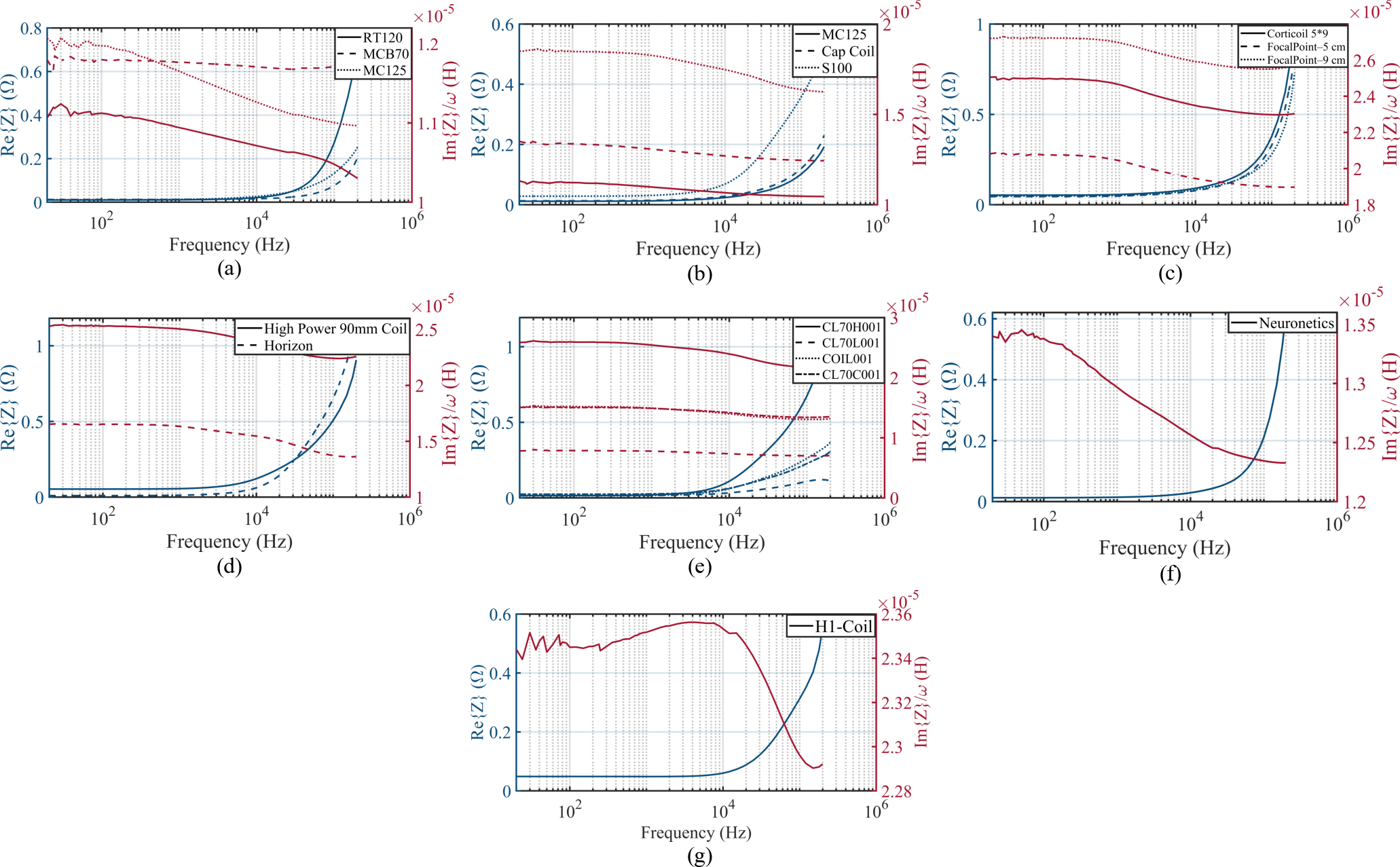} 
    \caption{Frequency-dependent coil impedance expressed as resistance and effective inductance. The real part of the impedance, $\Re\{Z(f)\}$, and the imaginary part normalised by angular frequency, $\Im\{Z(f)\}/\omega$, are shown for all coils and grouped by manufacturer: (a) MagVenture, (b) Dantec, (c) Cadwell, (d) Magstim, (e) Rogue Research, (f) Neuronetics, (g) BrainsWay.}
    \label{Bode_plots}
\end{figure*}
\end{landscape}

\begin{figure*}[!t]
    \centering
    \includegraphics[width=0.99\textwidth]{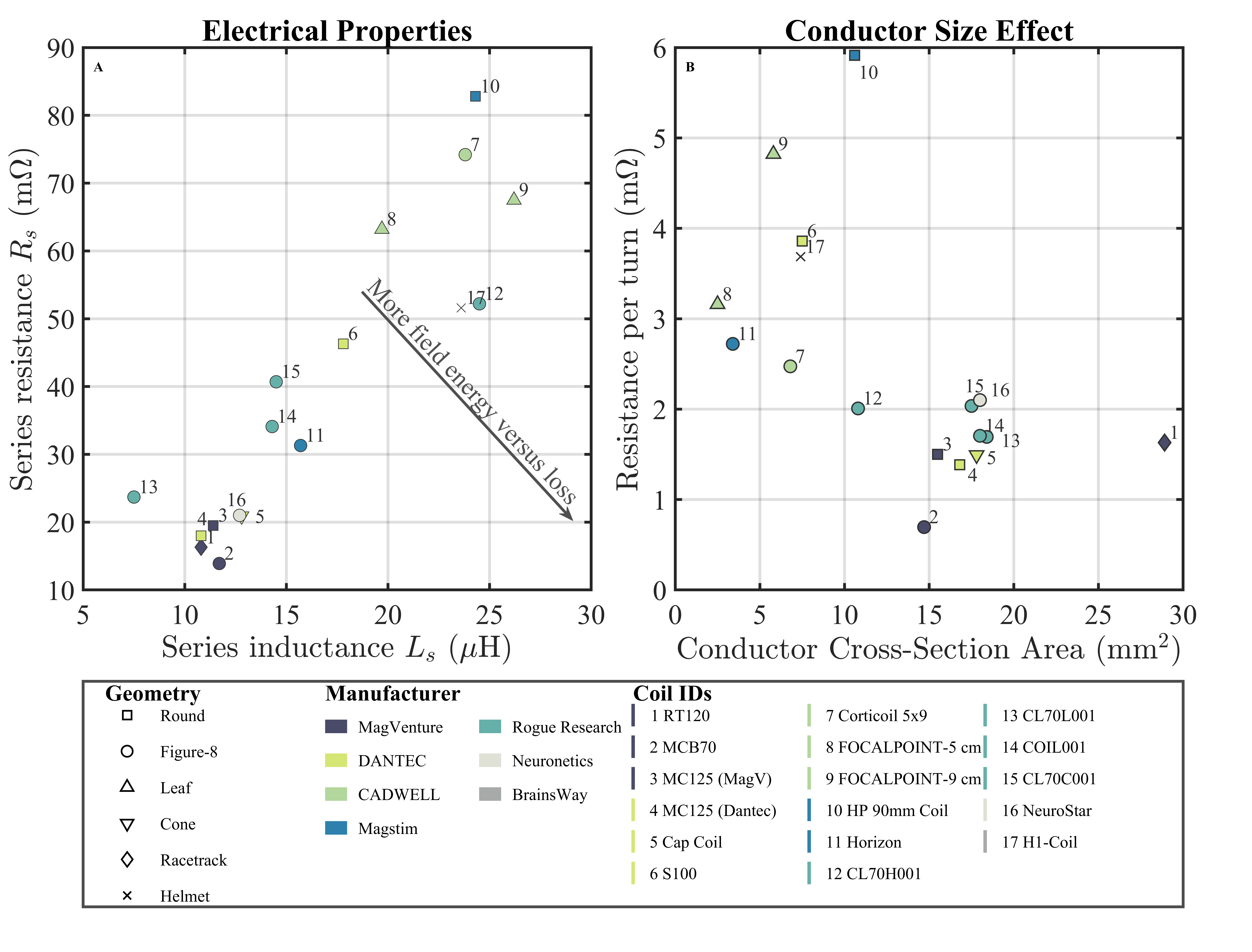} 
    \caption{Relationships between electrical measurements and conductor geometry across the surveyed TMS coils. (A) Series inductance $L_\mathrm{s}$ versus series resistance $R_\mathrm{s}$ at 5~kHz; (B) Resistance per turn $R_{\mathrm{per\ turn}}$ versus conductor cross-sectional area, showing an inverse trend in which larger conductors generally yield lower per-turn resistance.}
    \label{coil_analysis}
\end{figure*}

Figure~\ref{Bode_plots} presents the frequency-dependent real part of the coil impedance, \(\Re\{Z(f)\}\), and the effective inductance, \(\Im\{Z(f)\}/\omega\), for all surveyed coils. The curves provide a broadband electrical signature that complements the 5~kHz series parameters in Table~\ref{Coil_Electrical_Properties} and highlights coil-specific frequency-dependent behaviour.

\begin{landscape}
\begin{table*}[!t]
\centering
\caption{Conductor design parameters in the surveyed TMS coils.}
\label{Coil_CrossSection}
\footnotesize
\setlength{\tabcolsep}{5pt}
\renewcommand{\arraystretch}{1.12}

\begin{tabular}{@{} c l p{3.8cm} l c
S[table-format=2.2]
S[table-format=2.1]
c @{}}
\toprule
\textbf{Nr} & \textbf{Manufacturer} & \textbf{Model} &
\textbf{Conductor} & \textbf{Turns} &
\multicolumn{1}{c}{\textbf{Conductor Cross-}} &
\multicolumn{1}{c}{\textbf{Winding}} &
\textbf{Conductor} \\

& & & \multicolumn{1}{c}{\textbf{type}} & &
\multicolumn{1}{c}{\textbf{Sectional Area (mm$^{2}$)}} &
\multicolumn{1}{c}{\textbf{height (mm)}} &
\textbf{length (m)} \\
\midrule
1  & MagVenture     & RT120                  & Litz wire   & 10             & 28.9 & 14   & 3.5 \\
2  & MagVenture     & MCB70                  & Solid bar & 2$\times$10    & 14.7 & 5.6  & 4.1 \\
3  & MagVenture     & MC125                  & Litz wire & 13             & 15.5 & 5.9  & 3.2 \\
4  & Dantec         & MC125                  & Litz wire & 13             & 16.8 & 5.5  & 3.1 \\
5  & Dantec         & Cap Coil               & Solid bar & 14             & 17.8 & 5.7  & 3.5 \\
6  & Dantec         & S100                   & Litz wire   & 12             & 7.5  & 7    & 3.3 \\
7  & Cadwell        & Corticoil 5$\times$9   & Litz wire   & 2$\times$15    & 6.8  & 11   & 4.2 \\
8  & Cadwell        & FocalPoint--5 cm       & Litz wire   & 20             & 2.5  & 10.9 & 2.8 \\
9  & Cadwell        & FocalPoint--9 cm       & Litz wire   & 14             & 5.8  & 11   & 3.9 \\
10 & Magstim        & High Power 90 mm Coil  & Solid bar & 14             & 10.6 & 5.9  & 4.3 \\
11 & Magstim        & Horizon 3.0 E-z Cool   & Litz wire   & 6$\times$23             & 3.5  & 10.6 & 29.6 \\
12 & Rogue Research & CL70H001               & Litz wire   & 2$\times$13    & 10.8 & 11   & 5.8 \\
13 & Rogue Research & CL70L001               & Litz wire   & 2$\times$7     & 18.4 & 11   & 3.2 \\
14 & Rogue Research & COIL001                & Litz wire   & 2$\times$10    & 18   & 11   & 4.6 \\
15 & Rogue Research & CL70C001               & Litz wire   & 2$\times$10    & 17.5 & 11   & 4.7 \\
16 & Neuronetics    & NeuroStar              & Litz wire   & 2$\times$5     & 18   & 12.2 & 2.2 \\
17 & BrainsWay      & H1 Coil                & Litz wire & 14             & 7.42 & 3    & 14.7 \\
\bottomrule
\end{tabular}
\end{table*}
\end{landscape}

\begin{landscape}
\begin{table*}[!t]
\centering
\caption{Electrical properties of TMS coils: series inductance and resistance at 5 kHz, and dc resistance.}
\label{Coil_Electrical_Properties}
\footnotesize
\setlength{\tabcolsep}{6pt}
\renewcommand{\arraystretch}{1.15}

\begin{tabular}{@{} c l p{4.4cm} S[table-format=2.1] S[table-format=2.1] S[table-format=2.1] @{}}
\toprule
\textbf{Nr} & \textbf{Manufacturer} & \textbf{Model} &
\multicolumn{1}{c}{\textbf{$L_\mathrm{s}$ at 5 kHz (\si{\micro\henry})}} &
\multicolumn{1}{c}{\textbf{$R_\mathrm{s}$ at 5 kHz (\si{\milli\ohm})}} &
\multicolumn{1}{c}{\textbf{$R_\mathrm{DC}$ (\si{\milli\ohm})}}\\
\midrule
1  & MagVenture      & RT120                 & 10.8 & 16.3 & 10 \\
2  & MagVenture      & MCB70                 & 11.7 & 13.9 & 10 \\
3  & MagVenture      & MC125                 & 11.4 & 19.5 & 13 \\
4  & Dantec          & MC125                 & 10.8 & 18.0 & 15 \\
5  & Dantec          & Cap Coil              & 12.8 & 20.9 & 17 \\
6  & Dantec          & S100                  & 17.8 & 46.3 & 23 \\
7  & Cadwell         & Corticoil $5\times 9$ & 23.8 & 74.2 & 51 \\
8  & Cadwell         & FocalPoint--5 cm      & 19.7 & 63.2 & 44 \\
9  & Cadwell         & FocalPoint--9 cm      & 26.2 & 67.5 & 44 \\
10 & Magstim         & High Power 90 mm Coil         & 24.3 & 82.8 & 59 \\
11 & Magstim         & Horizon 3.0 E-z Cool  & 15.7 & 31.3 & 15 \\
12 & Rogue Research  & CL70H001              & 24.5 & 52.2 & 15 \\
13 & Rogue Research  & CL70L001              &  7.5 & 23.7 & 14 \\
14 & Rogue Research  & COIL001               & 14.3 & 34.1 & 13 \\
15 & Rogue Research  & CL70C001              & 14.5 & 40.7 & 22 \\
16 & Neuronetics     & NeuroStar             & 12.7 & 21.0 & 11 \\
17 & BrainsWay       & H1 Coil               & 23.6 &  51.6 & 46 \\
\bottomrule
\end{tabular}
\end{table*}
\end{landscape}

\subsection{Special Construction Features and Smart Coil Technologies}
Beyond loop geometry and conductor size, many coils also include additional construction features that support safety and usability. Table~\ref{Special_features} summarises noteworthy features.

Four coils (Magstim Horizon 3.0 E-z Cool, Rogue Research CL70C001, Neuronetics NeuroStar, and BrainsWay H1 Coil) use active air cooling.
Among the actively cooled designs, the Horizon 3.0 E-z Cool uses a ducted air channel with three axial fans in the handle that guides the air to underneath the winding. The coil  has fan-speed control. The air film below the winding increases the winding-to-cortex distance significantly, though.
The NeuroStar treatment coil also uses an integrated fan in conjunction with a temperature sensor to support active cooling and thermal monitoring.

Two coils incorporate a bent or angled surface to improve contact with a curved anatomy. To match the scalp curve, the MagVenture figure-of-eight design (MCB70-type) uses a well-known angled surface and a head angle of 150$^\circ$.
The NeuroStar coil similarly uses an angled contact face ($\sim$140$^\circ$) for a shorter gap for the magnetic flux outside the magnetic core to facilitate precise placement across the coil face.

Among the figure-of-eight coils, the Cadwell Corticoil 5$\times$9 has the smallest size and good focality. 

The BrainsWay H1 coil is integrated into a cushioned helmet that incorporates a cooling mechanism to help manage coil temperature during stimulation. In contrast to all other coils, the H1 coil is not potted into a rigid block. Only the inertia of the windings holds them in place against the forces during pulses. The system incorporates an efficient cooling mechanism that continuously streams air to the helmet to help maintain ambient coil temperature during repeated operation and to improve patient comfort. The analysed helmet used for depression studies contains both the active H1 coil and an integrated sham coil. The sham coil is designed to reproduce the scalp sensations and acoustic artefacts of active stimulation but avoids meaningful neuronal activation. Most sham elements are positioned well above the head, so the induced electric field in the brain may be negligible and insufficient to trigger neural activity.

\subsection{Electric-Field Simulation}

We simulated the CT-derived MCB70 and the SimNIBS MCB70 model under matched head-model, placement, distance, conductivity, and stimulation settings. Both simulations used the same \texttt{m2m\_ernie} head model, C3 target, CP3 coil orientation, input scalp distance, and stimulation rate of $\mathrm{d}I/\mathrm{d}t = 1.0\times10^{6}$ A/s. The final SimNIBS logs reported the same coil-placement matrix and the same coil-to-cortex distance of 18.59 mm for both models. Therefore, any observed differences reflect the effect of the coil-model representation under otherwise matched computational conditions.

\begin{figure}[!t]
    \centering
    \includegraphics[width=\columnwidth]{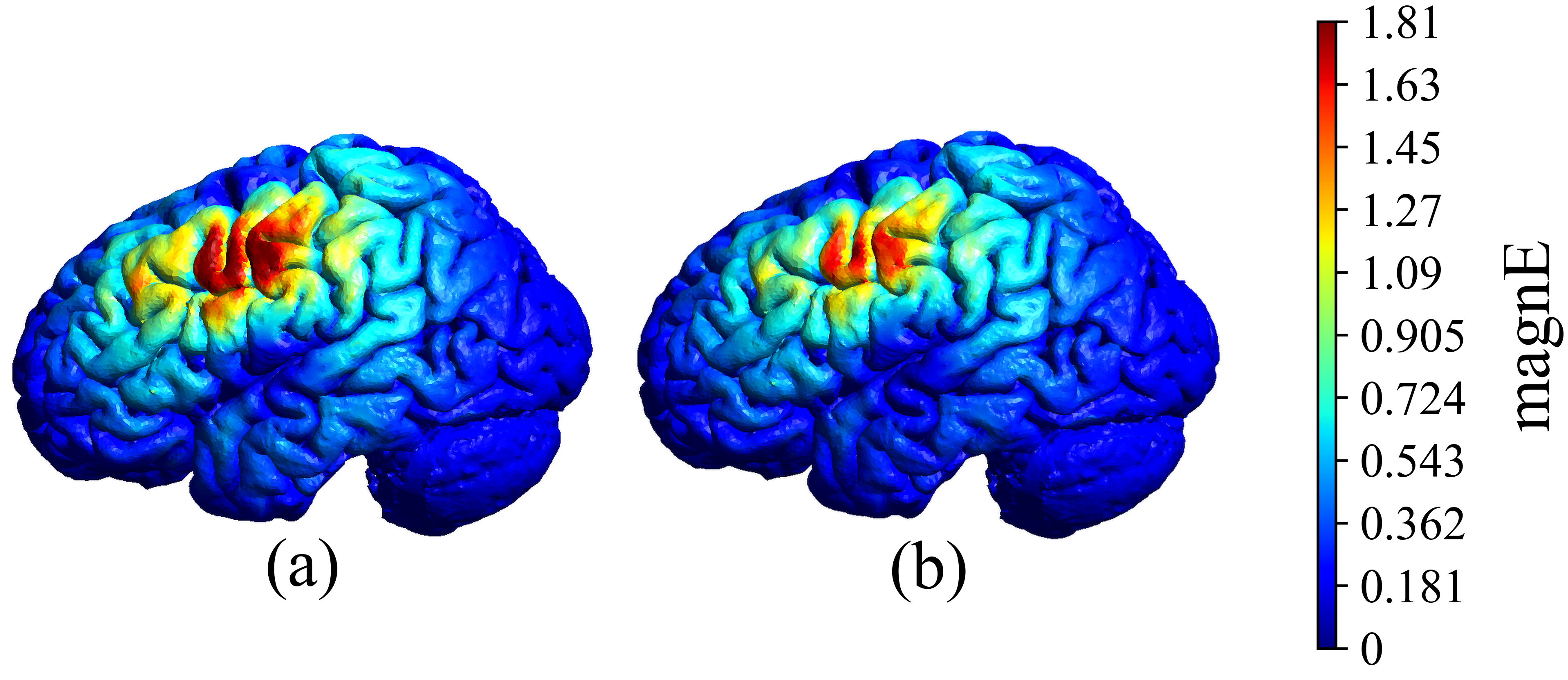}
 
    \caption{SimNIBS electric-field comparison between (a) the SimNIBS MCB70 model and (b) the CT-derived MCB70 coil model. Both simulations used identical head-model, target-location, coil-orientation, input-scalp-distance, stimulation-rate, coil-placement-matrix, and coil-to-cortex-distance settings. The target was C3, the coil y-direction pointed towards CP3, the stimulation rate was $\mathrm{d}I/\mathrm{d}t=1.0\times10^{6}$ A/s, and SimNIBS reported a coil-to-cortex distance of 18.59 mm for both models. The maps show the electric-field magnitude, \textit{magnE}, with the same colour scale.}
    \label{Efield_Comparison}
\end{figure}

Figure~\ref{Efield_Comparison} compares the resulting grey-matter electric-field distributions. The two models presented a similar overall spatial field pattern, as expected for the same figure-of-eight coil type and identical external placement. However, the CT-derived model yielded lower high-percentile electric-field magnitudes than the SimNIBS MCB70 model. Specifically, the CT-derived model gave $E_{99.9}=1.34$~V/m, $E_{99}=0.782$~V/m, and $E_{95}=0.399$~V/m, whereas the SimNIBS MCB70 model gave $E_{99.9}=1.64$~V/m, $E_{99}=0.947$~V/m, and $E_{95}=0.465$~V/m. The focality volumes remained similar between the two models: the CT-derived model gave $V_{75}=2.89\times10^{3}$~mm$^{3}$ and $V_{50}=1.20\times10^{4}$~mm$^{3}$, compared with $V_{75}=2.83\times10^{3}$~mm$^{3}$ and $V_{50}=1.15\times10^{4}$~mm$^{3}$ for the SimNIBS MCB70 model.

Relative to the SimNIBS MCB70 model, the CT-derived model produced lower high-percentile electric-field magnitudes, with differences of $-18.3\%$ for $E_{99.9}$, $-17.4\%$ for $E_{99}$, and $-14.2\%$ for $E_{95}$. The focality-volume differences were smaller, with $V_{75}$ increasing by $2.1\%$ and $V_{50}$ increasing by $4.3\%$. These results show that coil-model specification can affect quantitative electric-field metrics even when placement, distance, and stimulation settings remain constant.

\begin{table*}[!t]
\centering
\caption{Special construction features and smart technologies in the surveyed TMS coils.}
\label{Special_features}
\footnotesize
\setlength{\tabcolsep}{4pt}
\renewcommand{\arraystretch}{1.12}

\begin{tabularx}{\textwidth}{@{} c l p{3cm}  c c Y @{}}
\toprule
\textbf{Nr} & \textbf{Manufacturer} & \textbf{Model} &
\textbf{Cooling} &
\textbf{Notable internal features} \\
\midrule

1 & MagVenture & RT120  & Pasice       & -- \\ 
2 & MagVenture & MCB70  & Passice       & Bent/angled surface \\ 
3 & MagVenture & MC125  & Passive      & -- \\ 
4 & Dantec & MC125  & Passive    & -- \\ 
5 & Dantec & Cap Coil  & Passive      & Conical (cone) geometry \\ 
6 & Dantec & S100  & Passive     &  -- \\ 
7 & Cadwell & Corticoil 5$\times$9  &  Passive    & \makecell[c]{Compact 5,cm$\times$9,cm\\ design for high focality} \\
8 & Cadwell & FocalPoint–5 cm  & Passive   & -- \\ 
9 & Cadwell & FocalPoint–9 cm  & Passive  & -- \\ 
10 & Magstim & High Power 90 mm Coil  &  Passive    & -- \\  
11 &  Magstim & Horizon 3.0 E-z Cool  & Forced air & \makecell[c] {Multi-parallel windings} \\ 
12 & Rogue Research  & CL70H001  & Passive    & -- \\ 
13 & Rogue Research  & CL70L001  & Passive    & -- \\ 
14 & Rogue Research  & COIL001  & Passive    & -- \\ 
15 & Rogue Research  & CL70C001   &  Forced air &  3$\times$ internal fans \\ \\
16 & Neuronetics & NeuroStar    &  Forced air & \makecell[c] {Iron-core coil with\\ angled contact face} \\ \\
17 & BrainsWay      & H1 Coil    & Forced air & \makecell[c] {Helmet-integrated\\flexible H-coil, \\ implementation with\\integrated sham} \\
\bottomrule
\end{tabularx}
\end{table*}

\begin{table}[t]
\centering
\caption{Comparison of electric-field and focality metrics between the
CT-derived and SimNIBS MCB70 models for \(1\,\mathrm{A}/\mu\mathrm{s}\).}
\label{Simnibs_Comparison}

\footnotesize
\renewcommand{\arraystretch}{1.2}
\setlength{\tabcolsep}{4pt}

\begin{tabular*}{\columnwidth}{@{\extracolsep{\fill}}lccc@{}}
\toprule
\textbf{Metric} &
\textbf{CT-derived} &
\makecell{\textbf{SimNIBS}\\\textbf{MCB70 model}} &
\textbf{Difference} \\
\midrule
$E_{99.9}$ (V/m)   & 1.34                 & 1.64                 & $-18.3\%$ \\
$E_{99}$ (V/m)     & 0.782                & 0.947                & $-17.4\%$ \\
$E_{95}$ (V/m)     & 0.399                & 0.465                & $-14.2\%$ \\
$V_{75}$ (mm$^3$)  & $2.89\times10^{3}$   & $2.83\times10^{3}$   & $+2.1\%$  \\
$V_{50}$ (mm$^3$)  & $1.20\times10^{4}$   & $1.15\times10^{4}$   & $+4.3\%$  \\
\bottomrule
\end{tabular*}
\end{table}

\begin{figure*}[!t]
    \centering
    \includegraphics[width=\textwidth]{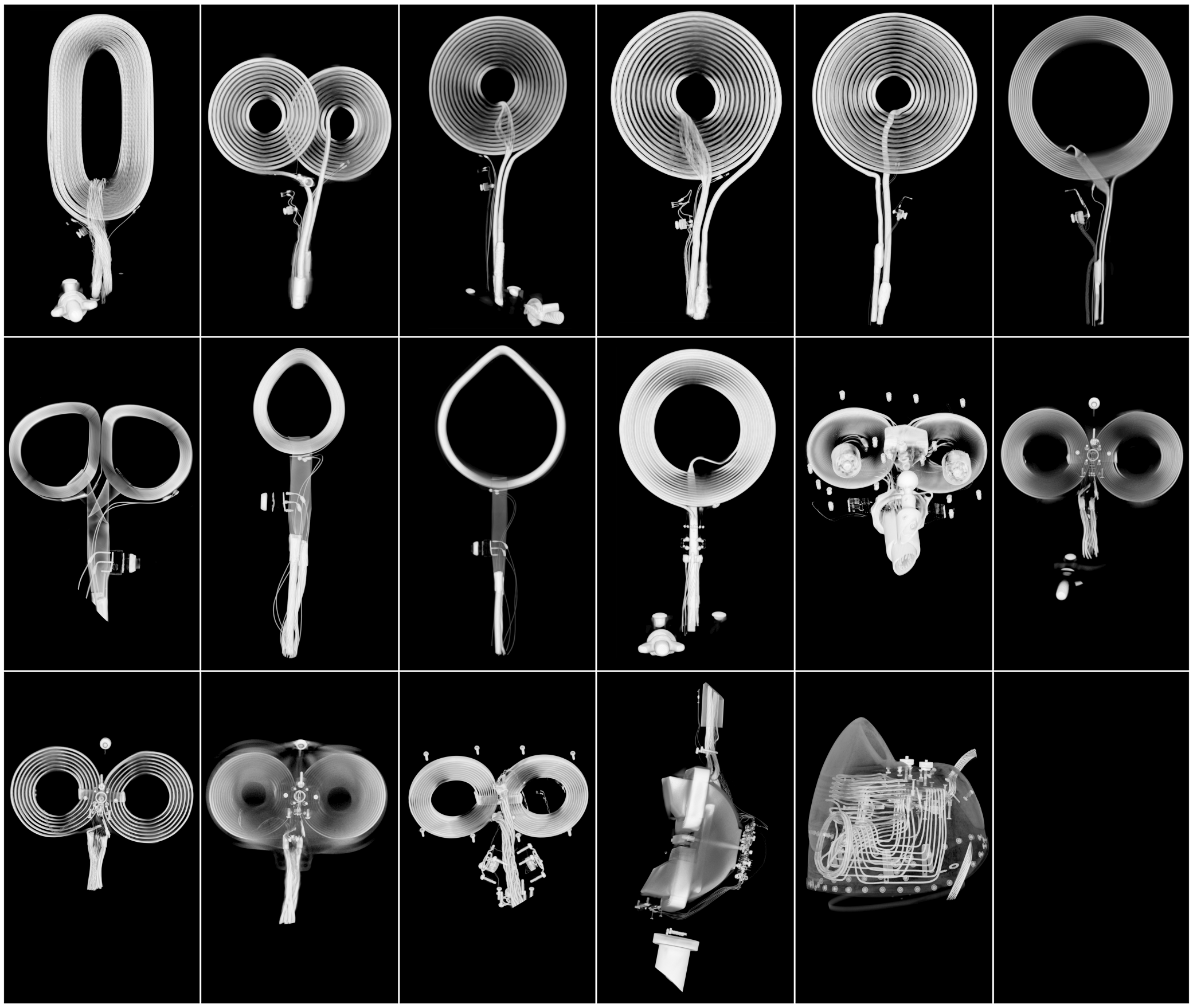} 
    \caption{High-resolution CT scan image, revealing the detailed internal structure of the coils. First row: MagVenture RT120, MagVenture MCB70, MagVenture MC125, Dantec MC125, Dantec Cap Coil, Dantec S100; second row: Cadwell Corticoil 5$\times$9, Cadwell FocalPoint--5 cm, Cadwell FocalPoint--9 cm, Magstim High Power 90 mm Coil, Magstim Horizon 3.0 E-z Cool, Rogue Research CL70H001; third row: Rogue Research CL70L001, Rogue Research COIL001, Rogue Research CL70C001, Neuronetics NeuroStar, BrainsWay H1 Coil.}
    \label{fig1}
\end{figure*}

\section{Discussion}

This study establishes a benchmark comparison across commercial TMS coils and provides coil-specific geometric and electrical parameters needed for accurate modelling. Figure~\ref{fig1} provides an overview of the micro-CT data across the surveyed coils and illustrates representative internal winding-pack and casing structures used for the geometric measurements reported in this paper. A key practical parameter is the distance from the patient contact surface to the internal windings. Across the tested coils, this coil–casing gap spans 1.7 to 7.5~mm. Many simulation studies assume a fixed 5~mm spacing, but this simplification can introduce meaningful errors in predicted electric fields. Tall winding packs add to that distance as they move the average current position farther away. Accurate field calculations therefore require coil-specific geometric inputs.

The SimNIBS simulation further demonstrates the modelling relevance of the CT-derived parameters. By calibrating the MCB70 line-current depth from the measured casing-to-winding distance and winding height, we constructed a coil-specific simulation model directly from the CT dataset. Under identical external placement conditions, this CT-derived model produced a spatially similar but quantitatively different electric-field distribution compared with the SimNIBS MCB70 model (Fig. \ref{Efield_Comparison}, Table \ref{Simnibs_Comparison}). These results support the practical utility of the dataset for constructing coil-specific simulation models and highlight the influence of internal winding geometry and winding-to-surface distance on predicted electric-field magnitudes.

Conductor technology also affects coil performance. Manufacturers use either solid-bar or litz-wire conductors to balance efficiency and heating. Litz-wire conductors cost more but offer lower high-frequency losses, which are already highly relevant in the frequency range of TMS pulses around 1 to 6~kHz.
Typical conductor cross-sections are in the range of 5 to 20~mm$^2$. A larger cross-sectional area helps increase the inductance-to-resistance ratio, which with proper coil design and low flux leakage increases the field for lower losses. MagVenture's MCB70 demonstrated the highest ratio in this category.

The surveyed coils incorporate different cooling technologies. Air-film cooling with air flow underneath the winding has to trade off the increased distance of the winding to the cortex, which in turn increases the current for the same stimulation field and therefore heating, which is linked to the squared current.  

Overall, these results support practical guidelines for coil modelling and fair comparison. Electric-field models should use coil-specific conductor-to-surface distances rather than a generic or otherwise assumed gap value, and they should incorporate turn count, conductor type, and cross-section. For a given stimulator, these parameters influence the delivered current waveform through their effect on $L_\mathrm{s}$ and $R_\mathrm{s}$.

\section{Conclusion}

This paper presents a technology benchmark and modelling-ready dataset for seventeen widely used TMS coils that  combines high-resolution micro-CT reconstructions with standardised electrical measurements. The micro-CT data resolve winding and housing structures and provide coil-specific geometric parameters relevant to numerical modelling, including loop dimensions, conductor cross-section, turn count, winding height, and winding-to-surface (casing) thickness. Complementary measurements report series resistance and inductance throughout the entire frequency range relevant to TMS. Together, these data enable reproducible coil-specific electric-field simulations, reduce uncertainty introduced by generic geometric assumptions (e.g., assumed coil--scalp spacing), and provide a quantitative reference for the design and evaluation of new coil technologies against established solutions. As a simulation-readiness example, we implemented the calibrated CT-derived MCB70 model in SimNIBS and demonstrated that the reported geometric parameters can be converted into coil-specific simulation models. The resulting electric-field estimates differed under otherwise identical placement conditions. Thus, the CT-derived coil geometry has the potential to improve the accuracy of coil-field modelling.

\section*{Data Availability}
The micro-CT dataset acquired in this study is publicly available through the RPTU Research Data Repository at https://doi.org/10.26204/data/30.

\bibliographystyle{elsarticle-num}
\bibliography{Manuscript}

\begin{thebibliography}{10}
\expandafter\ifx\csname url\endcsname\relax
  \def\url#1{\texttt{#1}}\fi
\expandafter\ifx\csname urlprefix\endcsname\relax\def\urlprefix{URL }\fi
\expandafter\ifx\csname href\endcsname\relax
  \def\href#1#2{#2} \def\path#1{#1}\fi

\bibitem{rossini2015non}
P.~M. Rossini, D.~Burke, R.~Chen, L.~G. Cohen, Z.~Daskalakis, R.~Di~Iorio, V.~Di~Lazzaro, F.~Ferreri, P.~Fitzgerald, M.~S. George, et~al., Non-invasive electrical and magnetic stimulation of the brain, spinal cord, roots and peripheral nerves: Basic principles and procedures for routine clinical and research application. an updated report from an ifcn committee, Clinical neurophysiology 126~(6) (2015) 1071--1107.

\bibitem{lu2017comparison}
M.~Lu, S.~Ueno, Comparison of the induced fields using different coil configurations during deep transcranial magnetic stimulation, PloS one 12~(6) (2017) e0178422.

\bibitem{krings1997stereotactic}
T.~Krings, B.~R. Buchbinder, W.~E. Butler, K.~H. Chiappa, H.~J. Jiang, B.~R. Rosen, G.~R. Cosgrove, Stereotactic transcranial magnetic stimulation: correlation with direct electrical cortical stimulation, Neurosurgery 41~(6) (1997) 1319--1326.

\bibitem{drakaki2022database}
M.~Drakaki, C.~Mathiesen, H.~R. Siebner, K.~Madsen, A.~Thielscher, Database of 25 validated coil models for electric field simulations for tms, Brain stimulation 15~(3) (2022) 697--706.

\bibitem{thielscher2004electric}
A.~Thielscher, T.~Kammer, Electric field properties of two commercial figure-8 coils in tms: calculation of focality and efficiency, Clinical neurophysiology 115~(7) (2004) 1697--1708.

\bibitem{tofts1990distribution}
P.~Tofts, The distribution of induced currents in magnetic stimulation of the nervous system, Physics in Medicine \& Biology 35~(8) (1990) 1119.

\bibitem{branston1991analysis}
N.~Branston, P.~Tofts, Analysis of the distribution of currents induced by a changing magnetic field in a volume conductor, Physics in medicine \& biology 36~(2) (1991) 161.

\bibitem{esselle2002neural}
K.~P. Esselle, M.~A. Stuchly, Neural stimulation with magnetic fields: analysis of induced electric fields, IEEE Transactions on Biomedical Engineering 39~(7) (2002) 693--700.

\bibitem{jalinous1991technical}
R.~Jalinous, Technical and practical aspects of magnetic nerve stimulation, Journal of Clinical Neurophysiology 8~(1) (1991) 10--25.

\bibitem{ravazzani1996magnetic}
P.~Ravazzani, J.~Ruohonen, F.~Grandori, G.~Tognola, Magnetic stimulation of the nervous system: induced electric field in unbounded, semi-infinite, spherical, and cylindrical media, Annals of biomedical engineering 24~(5) (1996) 606--616.

\bibitem{roth1990theoretical}
B.~J. Roth, L.~G. Cohen, M.~Hallett, W.~Friauf, P.~J. Basser, A theoretical calculation of the electric field induced by magnetic stimulation of a peripheral nerve, Muscle \& Nerve: Official Journal of the American Association of Electrodiagnostic Medicine 13~(8) (1990) 734--741.

\bibitem{ruohonen1998transcranial}
J.~Ruohonen, Transcranial magnetic stimulation: modelling and new techniques, Ph.D. thesis, J. Ruohonen (1998).

\bibitem{zimmermann1996slinky}
K.~P. Zimmermann, R.~K. Simpson, “slinky” coils for neuromagnetic stimulation, Electroencephalography and Clinical Neurophysiology/Electromyography and Motor Control 101~(2) (1996) 145--152.

\bibitem{eaton1992electric}
H.~Eaton, Electric field induced in a spherical volume conductor from arbitrary coils: application to magnetic stimulation and meg, Medical and Biological Engineering and Computing 30~(4) (1992) 433--440.

\bibitem{leo1992computer}
R.~D. Leo, G.~Cerri, D.~Balducci, F.~Moglie, O.~Scarpino, M.~Guidi, Computer modelling of brain cortex excitation by magnetic field pulses, Journal of medical engineering \& technology 16~(4) (1992) 149--156.

\bibitem{saypol1991theoretical}
J.~M. Saypol, B.~J. Roth, L.~G. Cohen, M.~Hallett, A theoretical comparison of electric and magnetic stimulation of the brain, Annals of biomedical engineering 19~(3) (1991) 317--328.

\bibitem{toschi2009transcranial}
N.~Toschi, T.~Welt, M.~Guerrisi, M.~E. Keck, Transcranial magnetic stimulation in heterogeneous brain tissue: clinical impact on focality, reproducibility and true sham stimulation, Journal of psychiatric research 43~(3) (2009) 255--264.

\bibitem{sekino2002comparison}
M.~Sekino, S.~Ueno, Comparison of current distributions in electroconvulsive therapy and transcranial magnetic stimulation, Journal of Applied Physics 91~(10) (2002) 8730--8732.

\bibitem{kowalski2002current}
T.~Kowalski, J.~Silny, H.~Buchner, Current density threshold for the stimulation of neurons in the motor cortex area, Bioelectromagnetics: Journal of the Bioelectromagnetics Society, The Society for Physical Regulation in Biology and Medicine, The European Bioelectromagnetics Association 23~(6) (2002) 421--428.

\bibitem{toschi2008reconstruction}
N.~Toschi, T.~Welt, M.~Guerrisi, M.~E. Keck, A reconstruction of the conductive phenomena elicited by transcranial magnetic stimulation in heterogeneous brain tissue, Physica medica 24~(2) (2008) 80--86.

\bibitem{sekino2004fem}
M.~Sekino, S.~Ueno, Fem-based determination of optimum current distribution in transcranial magnetic stimulation as an alternative to electroconvulsive therapy, IEEE Transactions on Magnetics 40~(4) (2004) 2167--2169.

\bibitem{salinas2007detailed}
F.~Salinas, J.~Lancaster, P.~Fox, Detailed 3d models of the induced electric field of transcranial magnetic stimulation coils, Physics in Medicine \& Biology 52~(10) (2007) 2879.

\bibitem{maccabee1993magnetic}
P.~Maccabee, V.~Amassian, L.~Eberle, R.~Cracco, Magnetic coil stimulation of straight and bent amphibian and mammalian peripheral nerve in vitro: locus of excitation., The Journal of physiology 460~(1) (1993) 201--219.

\bibitem{tay1989measurement}
G.~Tay, M.~Chilbert, J.~Battocletti, A.~Sances, T.~Swiontek, C.~Kurakami, Measurement of magnetically induced current density in saline in vivo, in: Images of the Twenty-First Century. Proceedings of the Annual International Engineering in Medicine and Biology Society,, IEEE, 1989, pp. 1167--1168.

\bibitem{wagner2004intracranial}
T.~Wagner, M.~Gangitano, R.~Romero, H.~Th{\'e}oret, M.~Kobayashi, D.~Anschel, J.~Ives, N.~Cuffin, D.~Schomer, A.~Pascual-Leone, Intracranial measurement of current densities induced by transcranial magnetic stimulation in the human brain, Neuroscience letters 354~(2) (2004) 91--94.

\bibitem{sances1991mapping}
A.~Sances~Jr, T.~Swiontek, Mapping of current densities induced in vivo during magnetic stimulation, in: Annual International Conference of the IEEEEngineering in Medicine and Biology Society, 1991.

\bibitem{lontis2006focality}
E.~R. Lontis, M.~Voigt, J.~J. Struijk, Focality assessment in transcranial magnetic stimulation with double and cone coils, Journal of Clinical Neurophysiology 23~(5) (2006) 463--472.

\bibitem{epstein2002iron}
C.~M. Epstein, K.~R. Davey, Iron-core coils for transcranial magnetic stimulation, Journal of Clinical Neurophysiology 19~(4) (2002) 376--381.

\bibitem{ueno2021figure}
S.~Ueno, M.~Sekino, Figure-eight coils for magnetic stimulation: from focal stimulation to deep stimulation, Frontiers in human neuroscience 15 (2021) 805971.

\bibitem{deng2014coil}
Z.-D. Deng, S.~H. Lisanby, A.~V. Peterchev, Coil design considerations for deep transcranial magnetic stimulation, Clinical Neurophysiology 125~(6) (2014) 1202--1212.

\bibitem{mancino2020obtaining}
A.~V. Mancino, F.~E. Milano, F.~M. Bertuzzi, C.~G. Yampolsky, L.~E. Ritacco, M.~R. Risk, Obtaining accurate and calibrated coil models for transcranial magnetic stimulation using magnetic field measurements, Medical \& Biological Engineering \& Computing 58~(7) (2020) 1499--1514.

\bibitem{bickford1965non}
R.~Bickford, B.~Freeming, Non-invasive magnetic stimulation of the human motor cortex, in: Digest of the 6th international conference of medical electronics in biology and engineering, 1965, p. 112.

\bibitem{polson1982stimulation}
M.~Polson, A.~Barker, I.~Freeston, Stimulation of nerve trunks with time-varying magnetic fields, Medical and Biological Engineering and Computing 20~(2) (1982) 243--244.

\bibitem{deng2017development}
S.~M. Goetz, Z.-D. Deng, The development and modelling of devices and paradigms for transcranial magnetic stimulation, International Review of Psychiatry 29~(2) (2017) 115--145.

\bibitem{ueno1990focal}
S.~Ueno, T.~Matsuda, M.~Fujiki, Focal stimulation of the human brain by a pair of opposing pulsed magnetic fields, International Journal of Applied Electromagnetics and Mechanics 1~(2-4) (1990) 167--172.

\bibitem{ueno1990functional}
S.~Ueno, T.~Matsuda, M.~Fujiki, Functional mapping of the human motor cortex obtained by focal and vectorial magnetic stimulation of the brain, IEEE Transactions on Magnetics 26~(5) (1990) 1539--1544.

\bibitem{koehler2024closed}
M.~Koehler, S.~M. Goetz, A closed formalism for anatomy-independent projection and optimization of magnetic stimulation coils on arbitrarily shaped surfaces, IEEE Transactions on Biomedical Engineering 71~(6) (2024) 1745--1755.

\bibitem{koponen2015minimum}
L.~M. Koponen, J.~O. Nieminen, R.~J. Ilmoniemi, Minimum-energy coils for transcranial magnetic stimulation: application to focal stimulation, Brain stimulation 8~(1) (2015) 124--134.

\bibitem{koehler2024coil}
M.~Koehler, T.~Kammer, S.~Goetz, How coil misalignment and mispositioning in transcranial magnetic stimulation affect the stimulation strength at the target., Clinical Neurophysiology 162 (2024) 159--161.

\bibitem{knaulein1996minimization}
R.~Kn{\"a}ulein, T.~Weyh, Minimization of energy stored in the magnetic field of air coils for medical application, International Workshop on electric and magnetic Fields 3 (1996) 477--482.

\bibitem{koponen2018multi}
L.~M. Koponen, J.~O. Nieminen, R.~J. Ilmoniemi, Multi-locus transcranial magnetic stimulation—theory and implementation, Brain Stimulation 11~(4) (2018) 849--855.

\bibitem{goetz2011comparison}
S.~Goetz, H.~Herzog, N.~Gattinger, B.~Gleich, Comparison of coil designs for peripheral magnetic muscle stimulation, Journal of neural engineering 8~(5) (2011) 056007.

\bibitem{goetz2022prediction}
S.~Goetz, J.~Kammermann, F.~Helling, T.~Weyh, Z.~Li, Prediction of force recruitment of neuromuscular magnetic stimulation from 3d field model of the thigh, IEEE Transactions on Neural Systems and Rehabilitation Engineering 30 (2022) 748--757.

\bibitem{kammer2003physikalische}
T.~Kammer, U.~Thielscher, {Physikalische und physiologische Grundlagen der transkraniellen Magnetstimulation}, Nervenheilkunde 22~(4) (2003) 168--76.

\bibitem{roth1994vitro}
B.~Roth, P.~Maccabee, L.~Eberle, V.~Amassian, M.~Hallett, J.~Cadwell, G.~Anselmi, G.~Tatarian, In vitro evaluation of a 4-leaf coil design for magnetic stimulation of peripheral nerve, Electroencephalography and Clinical Neurophysiology/Evoked Potentials Section 93~(1) (1994) 68--74.

\bibitem{lorenzen1992practical}
H.~Lorenzen, T.~Weyh, Practical application of the summation method for 3-d static magnetic field calculation of a setup of conductive and ferromagnetic material, IEEE transactions on magnetics 28~(2) (1992) 1481--1484.

\bibitem{al1998novel}
N.~Al-Mutawaly, R.~Findlay, A novel coil design for magnetic nerve stimulation, in: Conference Proceedings. IEEE Canadian Conference on Electrical and Computer Engineering (Cat. No. 98TH8341), Vol.~2, IEEE, 1998, pp. 669--672.

\bibitem{koehler2025magnetic}
M.~Koehler, A.~Sangle, S.~M. Goetz, Magnetic materials for transcranial magnetic stimulation ({TMS}), arXiv preprint arXiv:2511.00744 (2025).
\newblock \href {https://doi.org/10.48550/arXiv.2511.00744} {\path{doi:10.48550/arXiv.2511.00744}}.

\bibitem{roth2002coil}
Y.~Roth, A.~Zangen, M.~Hallett, A coil design for transcranial magnetic stimulation of deep brain regions, Journal of Clinical Neurophysiology 19~(4) (2002) 361--370.

\bibitem{roth2007three}
Y.~Roth, A.~Amir, Y.~Levkovitz, A.~Zangen, Three-dimensional distribution of the electric field induced in the brain by transcranial magnetic stimulation using figure-8 and deep h-coils, Journal of Clinical Neurophysiology 24~(1) (2007) 31--38.

\bibitem{huang2009consensus}
Y.-Z. Huang, M.~Sommer, G.~Thickbroom, M.~Hamada, A.~Pascual-Leonne, W.~Paulus, J.~Classen, A.~V. Peterchev, A.~Zangen, Y.~Ugawa, Consensus: New methodologies for brain stimulation, Brain stimulation 2~(1) (2009) 2--13.

\bibitem{harel2011h}
E.~V. Harel, A.~Zangen, Y.~Roth, I.~Reti, Y.~Braw, Y.~Levkovitz, H-coil repetitive transcranial magnetic stimulation for the treatment of bipolar depression: an add-on, safety and feasibility study, The World Journal of Biological Psychiatry 12~(2) (2011) 119--126.

\bibitem{koehler2024simplified}
M.~Koehler, S.~M. Goetz, Simplified h1 coil with a single layer of surface conductors, IEEE Access 12 (2024) 59861--59867.

\bibitem{filipvcic2019efficacy}
I.~Filip{\v{c}}i{\'c}, I.~{\v{S}}. Filip{\v{c}}i{\'c}, {\v{Z}}.~Milovac, S.~Su{\v{c}}i{\'c}, T.~Gaj{\v{s}}ak, E.~Ivezi{\'c}, S.~Ba{\v{s}}i{\'c}, {\v{Z}}.~Baji{\'c}, M.~Heilig, Efficacy of repetitive transcranial magnetic stimulation using a figure-8-coil or an h1-coil in treatment of major depressive disorder; a randomized clinical trial, Journal of psychiatric research 114 (2019) 113--119.

\bibitem{levkovitz2015efficacy}
Y.~Levkovitz, M.~Isserles, F.~Padberg, S.~H. Lisanby, A.~Bystritsky, G.~Xia, A.~Tendler, Z.~J. Daskalakis, J.~L. Winston, P.~Dannon, et~al., Efficacy and safety of deep transcranial magnetic stimulation for major depression: a prospective multicenter randomized controlled trial, World Psychiatry 14~(1) (2015) 64--73.

\bibitem{oreardon2007efficacy}
J.~P. O’Reardon, H.~B. Solvason, P.~G. Janicak, S.~Sampson, K.~E. Isenberg, Z.~Nahas, W.~M. McDonald, D.~Avery, P.~B. Fitzgerald, C.~Loo, et~al., Efficacy and safety of transcranial magnetic stimulation in the acute treatment of major depression: a multisite randomized controlled trial, Biological psychiatry 62~(11) (2007) 1208--1216.

\bibitem{lefaucheur2010image}
J.~Lefaucheur, Why image-guided navigation becomes essential in the practice of transcranial magnetic stimulation (2010).

\bibitem{goetz2021neuronavigation}
S.~M. Goetz, T.~Kammer, Neuronavigation, Oxford Handbook of Transcranial Stimulation (2021).

\bibitem{wang2023investigating}
Y.~Wang, Y.~Yang, Y.~Qi, E.~Gong, H.~Zhang, L.~Shi, Z.~Li, Investigating the electric field distribution in the human brain model induced by a high focality transcranial magnetic coil, AIP Advances 13~(10) (2023).

\bibitem{gomez2018design}
L.~J. Gomez, S.~M. Goetz, A.~V. Peterchev, Design of transcranial magnetic stimulation coils with optimal trade-off between depth, focality, and energy, Journal of neural engineering 15~(4) (2018) 046033.

\bibitem{stokes2005simple}
M.~G. Stokes, C.~D. Chambers, I.~C. Gould, T.~R. Henderson, N.~E. Janko, N.~B. Allen, J.~B. Mattingley, Simple metric for scaling motor threshold based on scalp-cortex distance: application to studies using transcranial magnetic stimulation, Journal of neurophysiology 94~(6) (2005) 4520--4527.

\bibitem{caulfield2024mitigating}
K.~A. Caulfield, S.~M. LaPorta, R.~M. Walton, E.~V. Collins, P.~M. Summers, J.~Y. Cho, M.~U. Antonucci, A.~Opitz, M.~S. George, L.~M. McTeague, Mitigating the risk of overdosing tms due to coil-to-scalp distance: An electric field modeling study, Brain Stimulation: Basic, Translational, and Clinical Research in Neuromodulation 17~(4) (2024) 970--974.

\bibitem{babbs2014compact}
C.~F. Babbs, A compact theory of magnetic nerve stimulation: predicting how to aim, BioMedical Engineering OnLine 13~(1) (2014) 53.

\bibitem{weyh2005marked}
T.~Weyh, K.~Wendicke, C.~Mentschel, H.~Zantow, H.~R. Siebner, Marked differences in the thermal characteristics of figure-of-eight shaped coils used for repetitive transcranial magnetic stimulation, Clinical neurophysiology 116~(6) (2005) 1477--1486.

\bibitem{elfahem2024advanced}
M.~Elfahem, Advanced modeling of litz wire using finite element method, Ph.D. thesis, {\'E}cole de technologie sup{\'e}rieure (2024).

\bibitem{gomez2020conditions}
L.~J. Gomez, M.~Dannhauer, L.~M. Koponen, A.~V. Peterchev, Conditions for numerically accurate tms electric field simulation, Brain stimulation 13~(1) (2020) 157--166.

\bibitem{koehler2023tms}
M.~K{\"o}hler, S.~Goetz, {TMS} coil design instrument (kl/codein box): A toolbox for creating user-defined coils from conductor path data, Brain Stimulation 16~(3) (2023) 698--700.

\bibitem{madariaga2007magnetic}
V.~B. Madariaga, A.~G. Manterola, E.~L. d. S.~M. Mir{\'o}, J.~B.~G. Iturri, Magnetic stimulation of the quadriceps: analysis of two stimulators used for diagnostic and therapeutic applications, Archivos de Bronconeumolog{\'\i}a (English Edition) 43~(7) (2007) 411--417.

\bibitem{szecsi2010force}
J.~Szecsi, S.~G{\"o}tz, W.~P{\"o}llmann, A.~Straube, Force--pain relationship in functional magnetic and electrical stimulation of subjects with paresis and preserved sensation, Clinical Neurophysiology 121~(9) (2010) 1589--1597.

\end{thebibliography}

\end{document}